\documentclass[aps,prl,amssymb,superscriptaddress,showpacs,twocolumn]{revtex4-1}

\usepackage{graphicx}
\usepackage{bm}% bold math
\usepackage{epsfig,amsmath, amsfonts, amssymb, xcolor,hyperref,wasysym}
\usepackage{ stmaryrd}
\usepackage{soul}
\usepackage{textcomp}

\begin{document}
\title{
Action at a distance. %
}
\author{D. B. Abraham}
\affiliation{Theoretical Physics, Department of Physics, University of Oxford,
1 Keble Road, Oxford OX1 3NP, United Kingdom}
\affiliation{Center for Non-linear Studies, Los Alamos National Laboratory, NM 87545, USA}

\author{A.~Macio\l ek}
\affiliation{Max-Planck-Institut f{\"u}r Intelligente Systeme, Heisenbergstr.~3, D-70569 Stuttgart, Germany}
\affiliation{IV. Institut f\"ur Theoretische  Physik, Pfaffenwaldring 57,  Universit\"at Stuttgart, D-70569 Stuttgart, Germany}

\affiliation{Institute of Physical Chemistry,
             Polish Academy of Sciences, Kasprzaka 44/52,
            PL-01-224 Warsaw, Poland}
\author{O. Vasilyev}
\affiliation{Max-Planck-Institut f{\"u}r Intelligente Systeme, Heisenbergstr.~3, D-70569 Stuttgart, Germany}
\affiliation{ IV. Institut f\"ur Theoretische  Physik, Pfaffenwaldring 57,  Universit\"at Stuttgart, D-70569 Stuttgart, Germany}

\date{\today}

\begin{abstract}

We present a system exhibiting 
giant proximity effects which parallel observations in superfluid helium  \cite{PKMG}
and   give a  theoretical explanation of these phenomena based on the mesoscopic picture of  phase coexistence
in finite systems. Our theory  is  confirmed by  MC simulation studies. 
Our work demonstrates  that such action-at-a-distance can occur in classical systems involving simple or 
complex fluids, such as  colloid-polymer mixtures, or ferromagnets.
\end{abstract}


\maketitle

  Can correlation effects in a fluid confined in big but finite compartments  linked by small openings, such as  shallow channels, 
occur over  distances much larger than the bulk correlation length?  
Recently, Gasparini and co-workers \cite{PKMG,fisher} have demonstrated such rather striking ``action-at-a-distance''
effects in a two-dimensional array of microscopic boxes filled with superfluid $^4$He  
and linked by either channels or a uniform 
film.
The measurements of several responses show that under certain conditions
these boxes can be strongly coupled to the  neighboring ones. 
What seems to be crucial in  this work is  the  size of boxes and  connectors and the vicinity of the critical point \cite{PG}.
Perron et al \cite{PG} suggested that action-at-a-distance effects might be a more general feature of  systems
with phase transitions than is usually supposed, 
a view which we confirm in 
this work for uniaxial classical ferromagnets and their analogs 
 (simple fluids or  binary mixtures in the lattice gas approximation, all belonging to the Ising model universality class
 of critical phenomena).
% Fisher \cite{fisher} has given an illuminating discussion of this field, pointing out the significance of his and Au Yang's work on layered 
% planar  Ising systems using exact solutions \cite{fisher_au_yang}.

 In the lattice gas picture the space occupied by the system is divided into boxes,  either vacant
or containing  a single molecule, unit maximal occupation  is achieved 
by a judicious choice of the box size.
The state of a box at position
$i$ with integer coordinates is labeled by a spin variable
$\sigma_{i}=\pm 1$. In the absence of bulk ordering field, a configuration $\{ \sigma \}$ of such (classical) spins has an energy 
\begin{equation}
\label{eq:1}
E(\{ \sigma \}) = - J \sum_{\langle {ij} \rangle}  \sigma_{i}  \sigma_{j}.
 \end{equation}
The sum $\langle {ij} \rangle$ is taken  over all nearest-neighbor pairs $ij$
of spins  and  $J$ is the coupling constant. Since the spins are assumed to be in thermal
equilibrium with a bath
  at temperature $T$,
  the probability of any spin configuration $\{ \sigma \}$
 is given by $p(\{ \sigma \})=Z^{-1} \exp(-\beta E(\{\sigma \}))$,  where $\beta = 1/k_BT$ ($k_B$  is the Boltzmann constant), 
and $Z$ is the normalization. The Helmholtz free energy from the formula $F=-(1/\beta)\ln Z$.
For dimensionality $d\ge 2$, it is known that such systems undergo phase transition to a low-temperature, 
magnetically ordered (dense) state \cite{O44}. For the square lattice, the critical value of $K=J\beta$  is
given by $K_{c} = (1/2)\ln(1+\sqrt{2}) \approx 0.440687$ \cite{O44}.
When $d=3$, various estimations are available \cite{RCW}; $K_{c}(d=3) \approx 0.2216544(3)\sim K_{c}(d=2)/2$.
Thus a $3d$ lattice orders more easily (higher critical temperature); this is a compatible with  Griffiths' correlation
inequalities \cite{griffiths}.
\begin{figure}[h!]
\begin{center}
 \includegraphics[width = 0.5\textwidth]{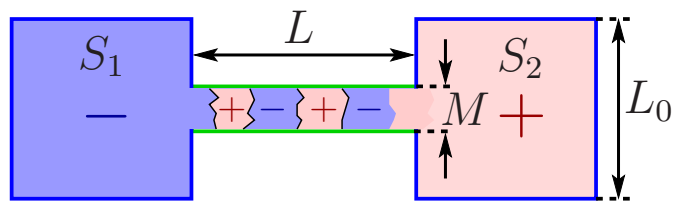}
 % geom.eps: 0x0 pixel, 300dpi, 0.00x0.00 cm, bb=0 -1 726 840
 \caption{(Color online) Side view  of an Ising system comprised of two cubic lattice boxes of a side $L_0$ 
 connected by a $L \times M, L \gg M$ strip. We assume  $ L_0 \gg M$. 
 \label{fig:1}
}
\end{center}
\end{figure}

Although an infinite system size is mandatory for a sharp  transition
\cite{O44}, Fisher and Privman \cite{privmann_fisher} gave a simple physical picture, denoted FP, of great elegance,
which captures the effect of 
finite size in a predictive way. Let us see how it works for a strip geometry in $d=2$ below the bulk critical temperature
$T_c$.
If we sum out fluctuations up to a certain length scale, a process often 
called coarse-graining
and take as a scale the bulk correlation length,
then a typical configuration 
in a strip is  one with 
regions of alternating $(+)$ and $(-)$  magnetization, with a magnitude roughly the spontaneous magnetization $m$,
separated by domain walls (Fig.~\ref{fig:1}).
The statistical weight of such a domain wall is  taken to  be $\tilde w = \exp(-M\tau)$,
where $M$ is the width of the strip and $\tau$ is the reduced (by a factor of $k_BT$) interfacial tension or a domain wall free energy. 
The configurational  entropy is estimated by treating the domain walls as point particles on 
the line which are strictly avoiding. Using this simple picture,  we can calculate exactly
 the pair correlation function $G(x)$ of two spins in
 the same edge of a strip  separated by a distance $x$ (we use throughout a lattice constant as a length unit):
\begin{equation}
\label{eq:2}
 G(x) \propto \frac{\left[(1+\tilde w)^x+(1-\tilde w)^x\right]-\left[(1+\tilde w)^x-(1-\tilde w)^x\right]}{2(1+\tilde w)^x}.
\end{equation}
The first term inside square brackets is the weight from  configurations with
an even number of domain walls between the spins;  evidently in this case the spins must be parallel.
The second term considers an odd number of such domain walls and is thus the contribution from anti-parallel spins.
Equation (\ref{eq:2}) should be complemented with  
the prefactor $m_e(M)$, which  becomes  the edge magnetization as $M\to\infty$ \cite{MacCoy_Wu}. 
Its value is a manifestation of fluctuations
on the scale of the bulk correlation length $\xi_b$ and thus is not accessible in FP. 
For sufficiently wide strips and fixed temperature below $T_c(d=2)$ such that $M \tau \gg 1$ ($\tilde w \ll 1$),  Eq.~(\ref{eq:2}) 
can be simplified to give a pure exponential decay: 
\begin{eqnarray}
 \label{eq:3}
 G(x)& = & m_e^2(M) \exp\left[ -x \log \left(\frac{1+\tilde w}{1-\tilde w}\right) \right] \nonumber \\ &&
 \sim m_e^2 \exp\left[-2x{\tilde w}(1+O({\tilde w}^2))\right].
\end{eqnarray}
It is noteworthy that in the FP picture the decay length of correlation function $G(x)$ {\it diverges exponentially} with $M$. 
We now  compare this prediction  with the result of exact calculation for  the full Ising strip   \cite{DBA}.
We find  agreement 
in the asymptotic behavior of $G(x)$ provided $\tilde w$ in
eq.~(\ref{eq:3}) is replaced  by
$ w =(\sinh 2K)^{-1}\sinh(\tau) e^{-M\tau}$.
This is easy to understand if we note that the simple Helmholtz fluctuation estimate expressed by $\tilde w$
must be modified to include the point tension (a $2d$ analogue of line tension) and this we calculate exactly \cite{AM} 
in confirmation. 
In the exact result of Ref.~\cite{DBA}, there is an additional contribution to $G(x)$
due to  fluctuations on the scale of the bulk correlation length $\xi_b$. Because in $2d$ Ising model
$\xi_b = 1/\tau$, this contribution is relatively negligible provided $w \ll \tau$. This gives us a criterion for the validity
of our theory which confirms na{\"i}ve expectations.
Stated another way,
the recapture of long range order  is achieved {\it not} through  bulk correlation length related phenomena, but rather  
by the emergence of a new length scale which  diverges exponentially
fast as $M\to \infty$, as $\exp(M\tau)$.
\begin{figure}[h!]
\begin{center}
 \includegraphics[width=0.5\textwidth]{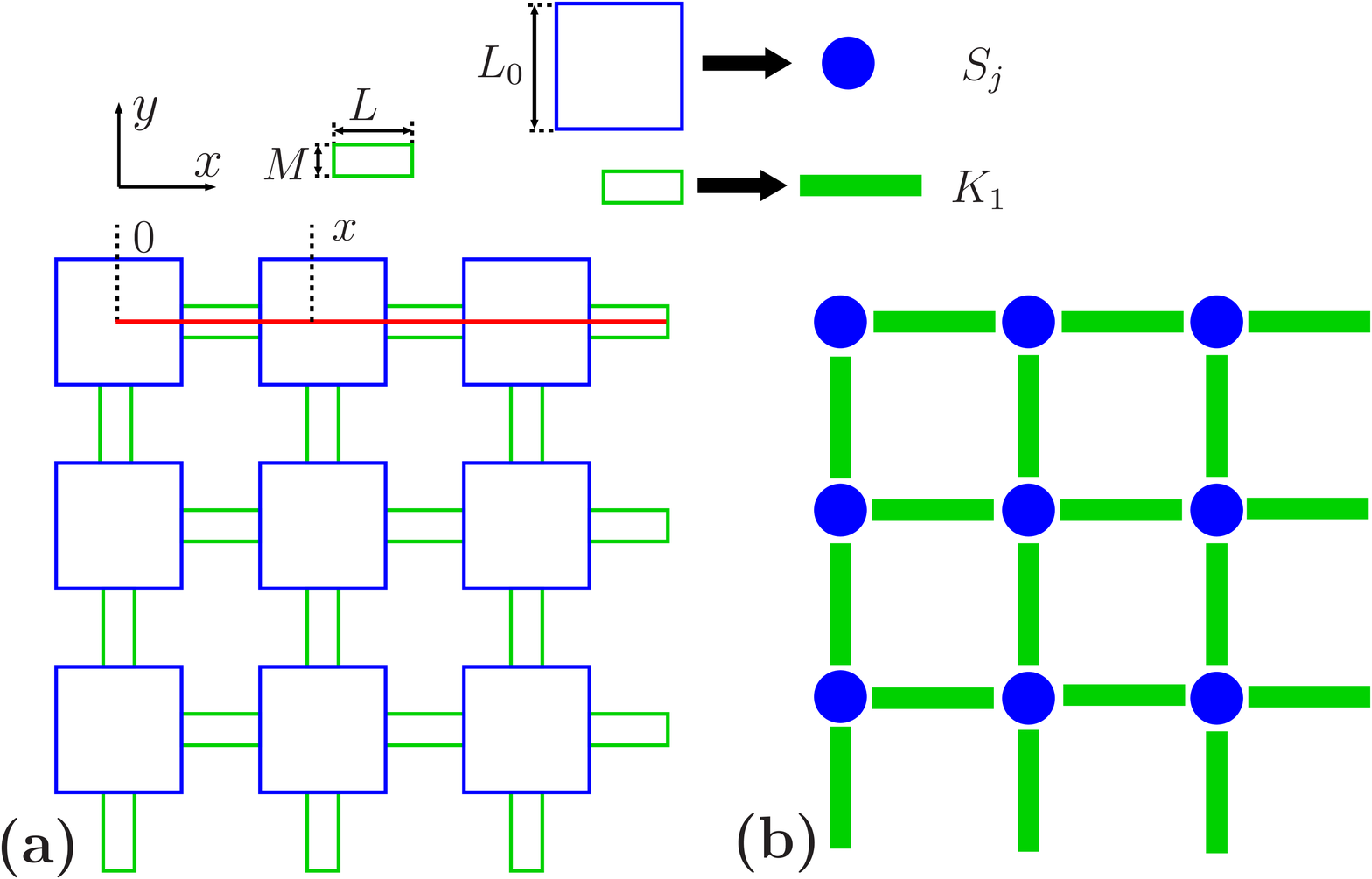}
 % geom.eps: 0x0 pixel, 300dpi, 0.00x0.00 cm, bb=0 -1 726 840
 \caption{(Color online)
(a)  Geometry of two-dimensional  array of  $N_{0}\times N_{0}$    cubes  of  size 
$L_{0}$ connected by strips (channels)
of length $L$ and thickness $M$. (b) The states of  boxes are  described by the spin variable $S_j = \pm 1$; the nearest-neighbor boxes
interact with the interaction energy $K_1S_iS_j$.
\label{fig:2}
}
\end{center}
\end{figure}

We now apply the FP idea with improved statistical   weight  $w$ of the domain wall, which we term  {\it enhanced} Fisher-Privman theory (EFP) \cite{EPL},
to the scheme of Fig.~\ref{fig:1}: the pair of cubic lattice boxes of side $L_0$ 
is coupled by an Ising strip of dimension $L \times M$, with $L_0 \gg M$.
For $T<T_c(d=2)$, the picture which emerges is one with  a sequence of domain walls crossing the strip,
but none inside the boxes, because they would be of 
the size $L_0 \times L_0$ and controlled by  a higher surface tension, thus  negligible. 
Because  $K_{c}(d=3)\sim K_{c}(d=2)/2$ and boxes  are large, we expect that 
the state of each box is either magnetized up or down 
and assignment can be described  on this level of coarse graining by a variable $S_j =\pm 1$ for each box, $j=1,2$.
The Boltzmann factor  for a given assignment in place of argument of the $S_j$
is thus 
\begin{equation}
 \label{eq:4}
 Z = Z_e^{(1+S_1S_2)/2}Z_o^{(1-S_1S_2))/2} = Ae^{K_1S_1S_2}
\end{equation}
 where $Z_o$ (resp. $Z_e$)  is the partition function for an odd (resp.even) number of domain walls and
the interaction constant  $K_1$ is given by  
\begin{equation}
 \label{eq:5}
  e^{2K_1}= \frac{Z_e}{Z_o} = \frac{1+t^L}{1-t^L}, \quad t=\frac{1-w}{1+w}.
\end{equation}
\begin{figure}[h!]
\begin{center}
 \includegraphics[width=0.45\textwidth]{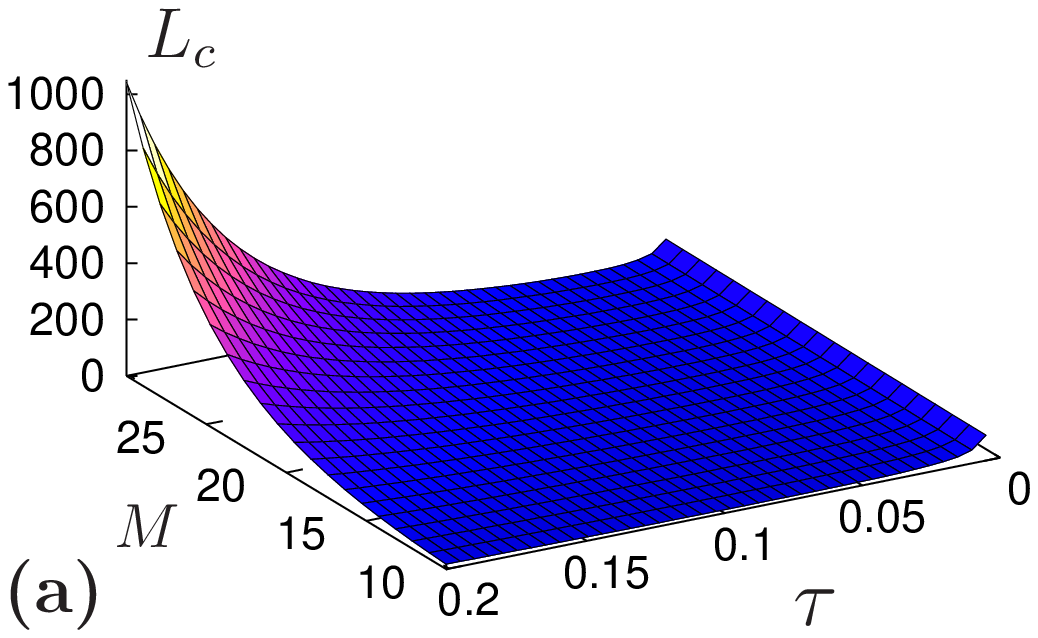}
 \includegraphics[width=0.4\textwidth]{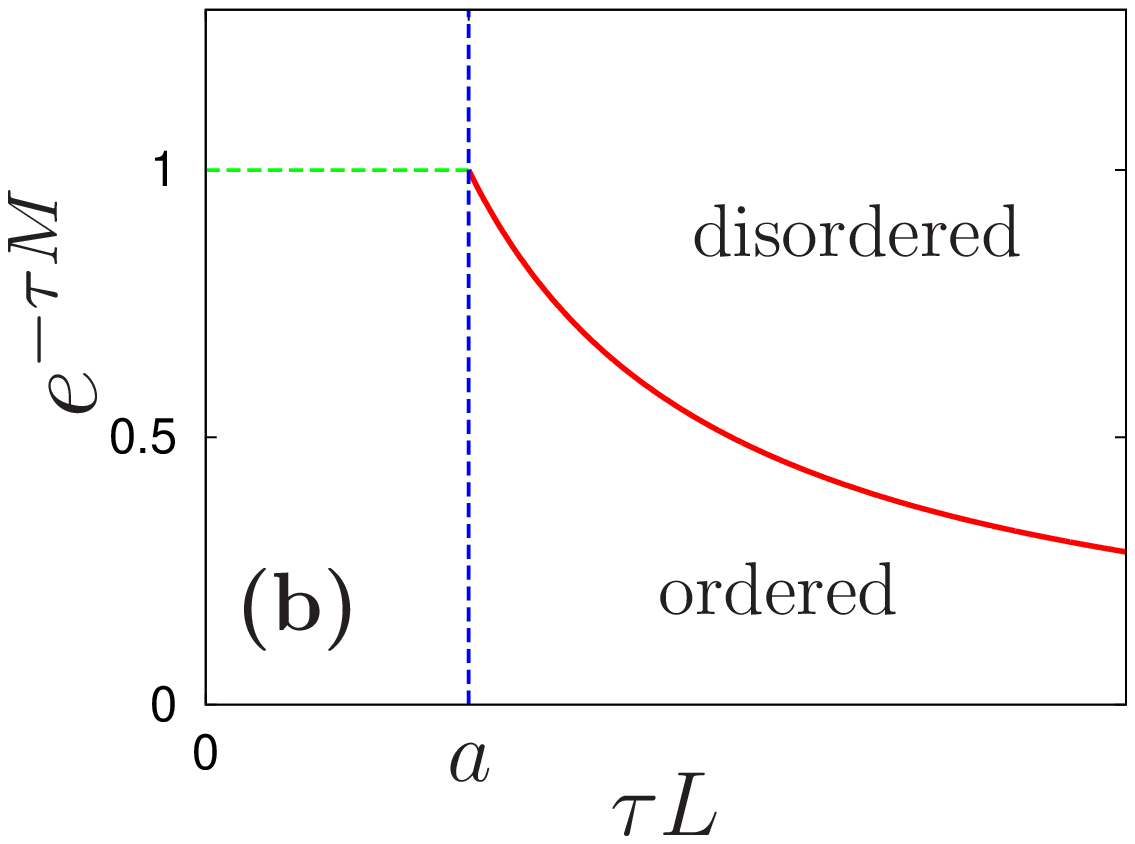}
 % geom.eps: 0x0 pixel, 300dpi, 0.00x0.00 cm, bb=0 -1 726 840
 \caption{(Color online)  The phase diagram of the two-dimensional ''network`` lattice shown in Fig.~\ref{fig:2}.
 (a) The critical value of the length of connecting strips $L_c$
  depends on the surface tension $\tau$
 and the width $M$ of the strip.
 Network is ordered in the region which lies below the critical surface $L_c(\tau,M)$.
 (b) The phase diagram in the scaling limit $M,L \to \infty$ and $\tau \sim T-T_c(d=2) \to 0$; $a=2^{-1}\ln(1+\sqrt 2)$.
\label{fig:3}
}
\end{center}
\end{figure}
We can  assemble such bonds, assumed mutually independent, and boxes to make up a ``network`` lattice.
For an analogous set up to that of Gasparini and co-workers we take a two dimensional array as illustrated in Fig.~\ref{fig:2}. 
The intriguing possibility is that $K_1$ could  satisfy $K_1 > K_{c}(d=2)$ 
by, e.g., adjusting the width or length of the strip 
at a fixed temperature $T<T_c(d=2)$.
Solving above with $K_1 = K_{c}(d=2)$  gives a critical value $L_c$ where
$L_c \ln(1/t(M,\tau)) = \ln (1 + \sqrt{2})$.
With $L< L_c(M,\tau)$, the system is subcritical and hence {\it ordered}. 
Because $w$ is small, this implies that $2L_cw = \ln (1+\sqrt{2})$; thus $L_c$ diverges as $\exp (M\tau)$.
In the EFP (contrary to FP), this result  scales near the $2d$ critical point.
Using scaling variables $ L_c\tau$ and $M\tau$ we obtain 
for the network critical point 
$L_c\tau e^{-M\tau} = 2^{-1}\ln (1 + \sqrt{2})$.
The critical surface   $L_c(M,\tau)$   shown in Fig.~\ref{fig:3}(a) displays an interesting feature:
for sufficiently wide strips, the critical value $L_c$ of  strip length is a non-monotonic function
of temperature. This offers a possibility to tune the collective behavior of  boxes
by varying the temperature. If the size of connecting strips is suitable chosen,
 the initially  correlated boxes become uncorrelated upon increasing $T$
 but then
correlated again sufficiently close to $T_c(d=2)$. 
\begin{figure}[h!]
\begin{center}
 \includegraphics[width=0.45\textwidth]{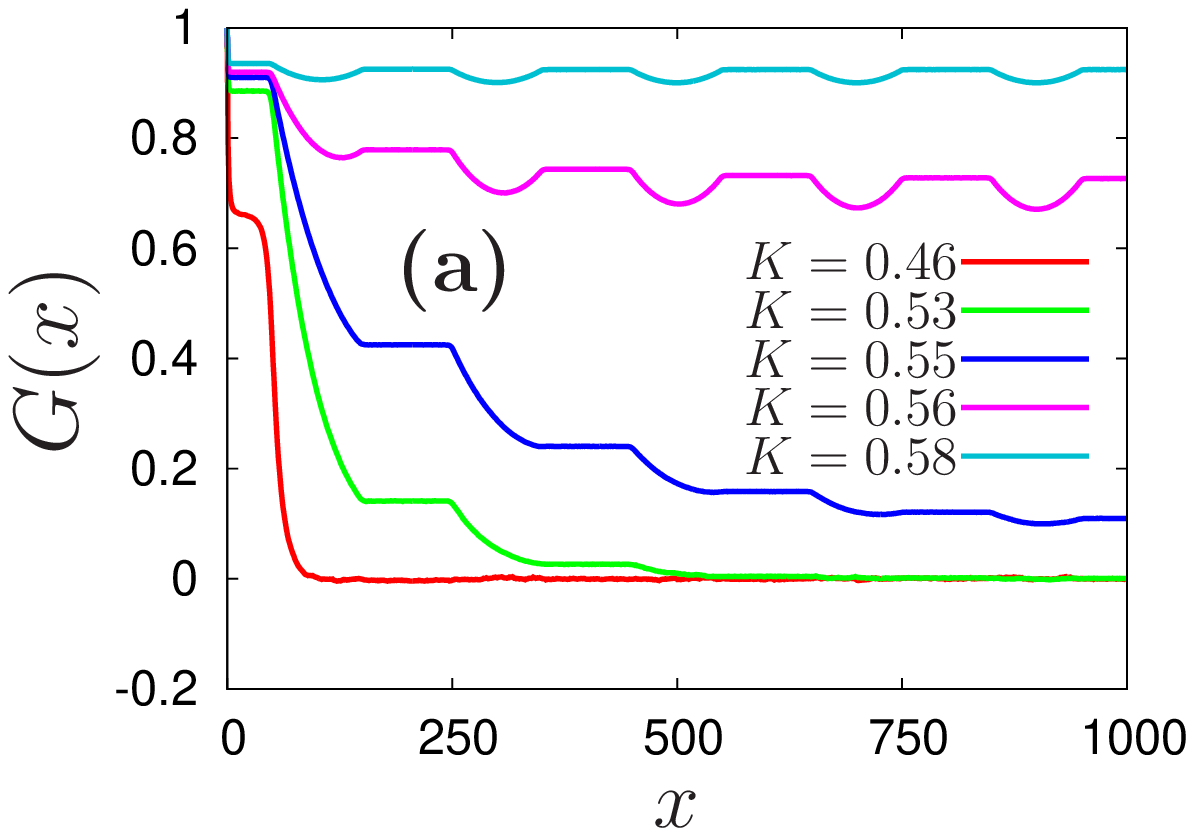}
 \includegraphics[width=0.45\textwidth]{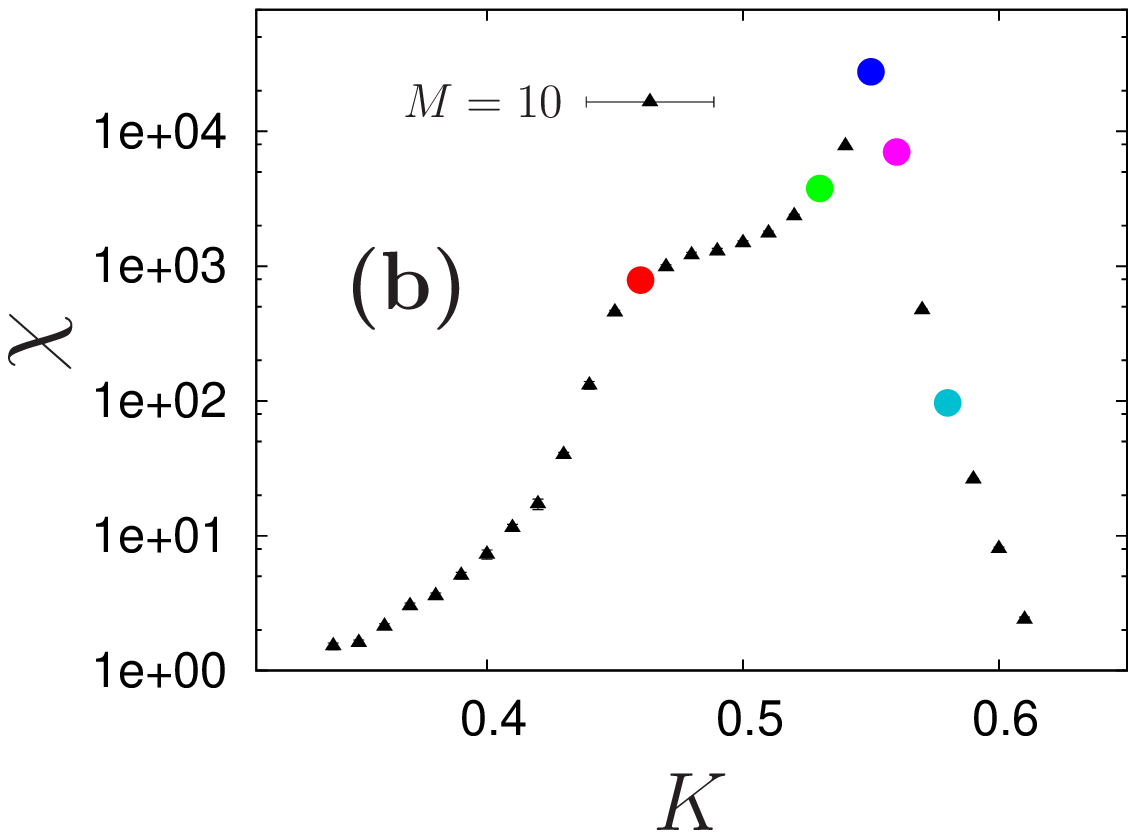}
 \caption{(Color online)   
 Monte Carlo simulation data for a $2d$ array of $10 \times 10$ squares with the side $L_{0}=100$
connected by strips of the length $L=100$ and the width 
$M=10$: (a) the spin-spin correlation function 
$G(x)=\langle \sigma(0)\sigma(x)\rangle$ as a function of the distance $x$ along the
center of the channel (red solid line in Fig~\ref{fig:2}) for various couplings $ K =0.46,0.53,0.55,0.56,0.58$ 
(results for smaller $K$ are not accessed from the EFP theory \cite{fisher_au_yang}) and
(b)  the  susceptibility $\chi$ of the system as a function of $K$. 
The values of $\chi$ at temperatures for which the correlation functions are shown in (a)
 are highlighted by points with the same color
as the corresponding curves $G(x)$. 
\label{fig:4}
}
\end{center}
\end{figure}

We have tested the application of EFP theory by MC simulation. 
The averaging has been performed over
$10^{4}-10^{5}$ of Monte Carlo steps. Each hybrid MC step consists 
of a flip of Wolff  cluster and application of Metropolis
updates to a randomly chosen quarter  of all spins in the system~\cite{LB}. 
For a $1d$ array of $2d$ Ising boxes and strips
we observe (results are not presented here)
that the pair  correlation
function below $T_c(d=2)$  has plateaux, which confirms our expectation that boxes are ordered. 
As predicted, these plateaux  values  approach zero following the Ising correlation function law  
$G(\tilde x) \simeq m^2(\tanh K_1)^{| x/L|}$ with  $\tanh K_1 = t^L$ where  $m$ is the (spontaneous) magnetization in the box. 
We have also performed MC simulation  of a $2d$ array of squares. We find that,  below  certain temperature, 
the pair correlation function 
$G(x)$  along the line connecting   centers
of squares via the  channels (red line in Fig.~\ref{fig:2}) does not decay to zero. This is 
a clear manifestation of the existence of
order in the network. For example, the system for which the MC data  are shown    
in Fig~\ref{fig:4}, undergoes the (rounded in the finite system) ordering transition 
at $K \approx 0.55$, which agrees perfectly  with the prediction from the EFP theory.
This transition lies {\it below} the critical point of the $2d$ Ising model, 
as implied by Griffiths’ inequalities; this is because such a lattice has been perforated to arrive at the network model.
It will be accompanied by a divergent susceptibility; in the  MC simulation data shown in Fig~\ref{fig:4}(b)
one can see a peak located at $K_m \simeq 0.55$. The ghost of the “rounded” phase transition in the connecting strips 
appears as a “shoulder” 
(red dot in Fig.~\ref{fig:4}(b)). This is analogous to the findings reported in \cite{fisher} and in \cite{fisher_au_yang}.
\begin{figure}[h!]
 \includegraphics[width=0.45\textwidth]{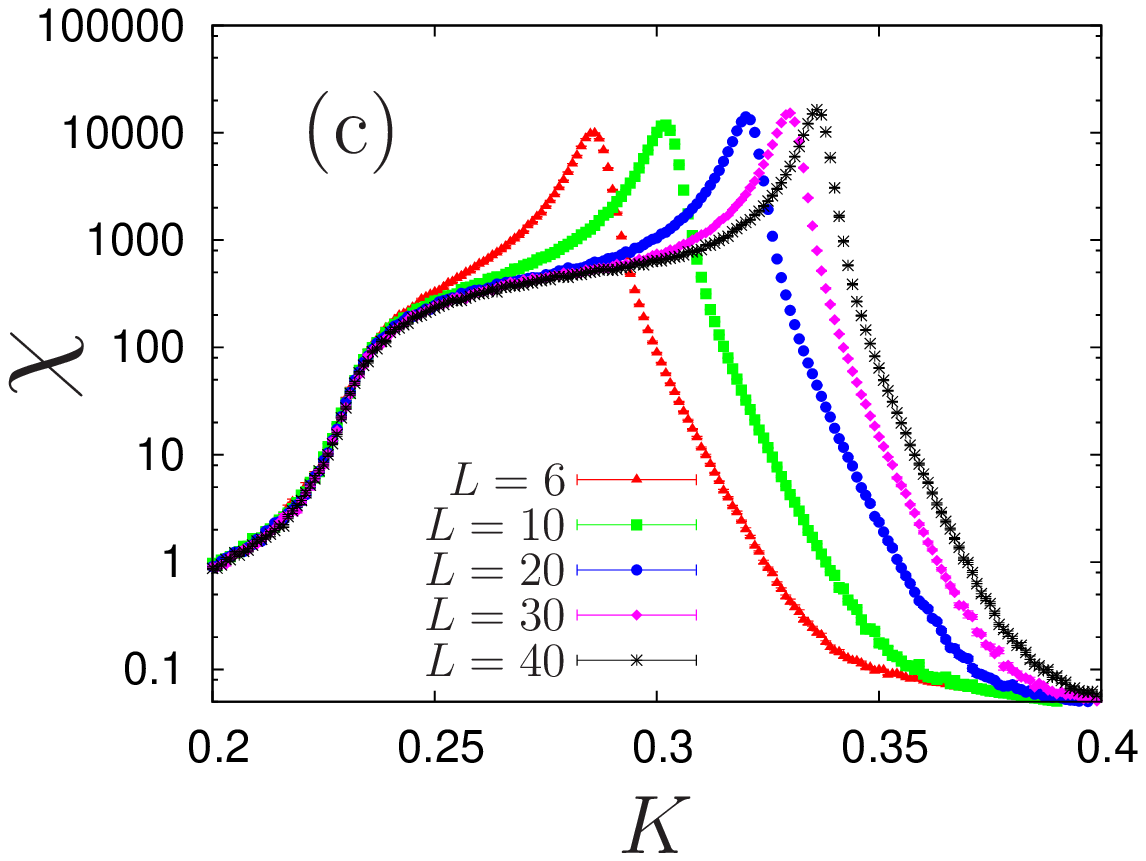}
 \includegraphics[width=0.45\textwidth]{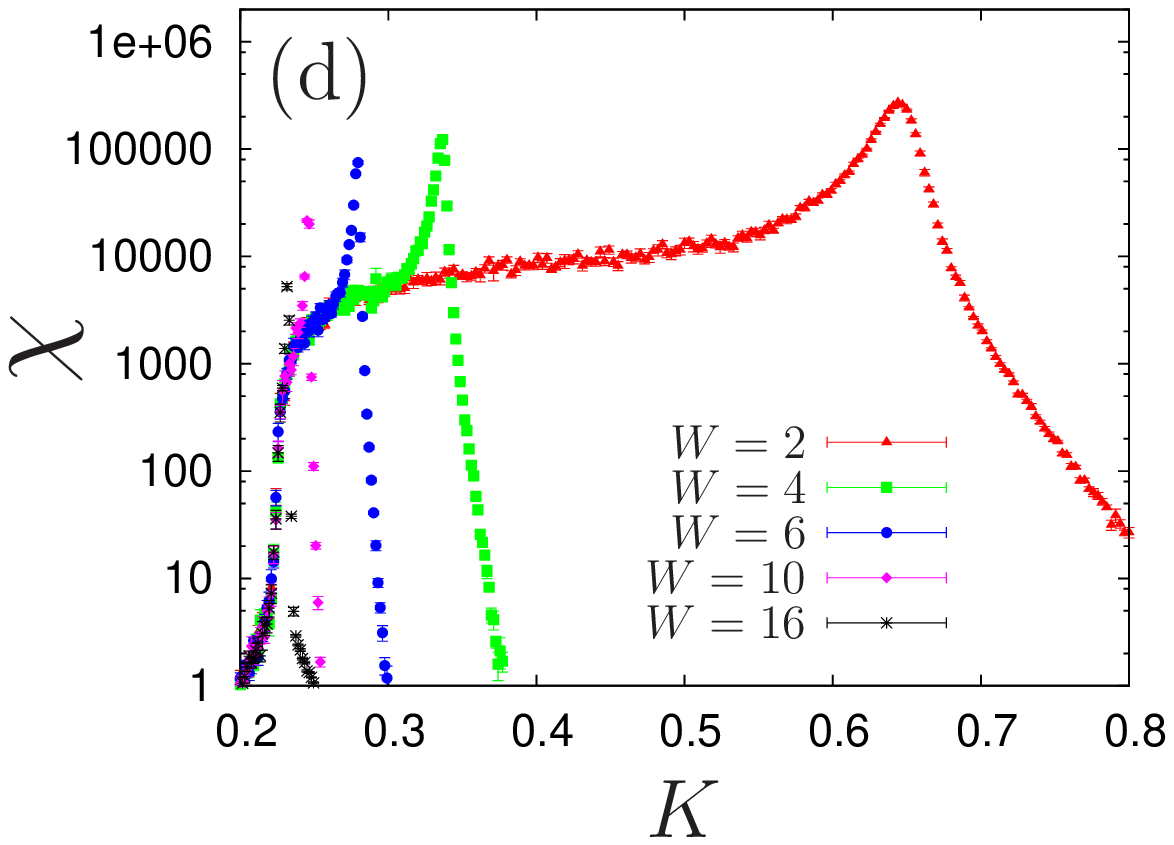}
 \caption{(Color online) 
 Magnetic susceptibility of  the  two-dimensional array of 
 $N_{0}\times N_{0}$ cubes of size $L_{0}^{3}$,
 connected by channels of size $L \times W\times W$
 as function of $K$ for:
(a) $N_{0}=10$, $L_{0}=20$, $W=4$  and  various $L=6,10,20,30,40$;
(b) $N_{0}=10$, $L_{0}=40$, $L=40$ and  various $W=2,4,6,10,16$.
 } 
\label{fig:5}
\end{figure}
Our EFP	arguments also apply to	 boxes connected by rods. But in this case there are 
no exact results available for the quantity $w$. We must resort	 to simulation	 for its 
evaluation, as	will be	 detailed elsewhere \cite{AMV}. The long-­ranged coupling between boxes should occur, much as in the
results reported above.  To confirm this anticipation, we have performed MC
simulation of  the  two-dimensional array 
of  $N_{0}\times N_{0}$   cubes of size $L_{0}^{3}$
 connected by channels of the length $L$
and cross-section $W\times W$ (the top view of this system
is shown in Fig.~\ref{fig:2}(a)). 
The magnetic susceptibility $\chi$ of such a   system
as  function of $K$ displays  features similar to these 
  of  the  $2d$ array of squares connected by strips.
We observe, as expected, that the only effect of
the system size $N_{0}$ on $\chi$ is to increase the height of the
peak at the ordering transition point.  
The cube size $L_{0}$  affects the shoulder of the curve  which is the remnant of the  $3d$ bulk transition,
 but does not change the location of the  peak.
The location of the ordering transition 
point is determined entirely  by the channel geometry $L$ and $W$,  
see Fig.~\ref{fig:5}(a),(b).

We have introduced in this work a theory supporting the intriguing suggestion by Perron et al \cite{PKMG} that 
the action at a distance effect which they observed experimentally in superfluid $^4$He might be a {\it widespread} consequence
of phase transitions and critical phenomena. Our theory, which applies to classical lattice gases and
their analogues, has a key ingredient: the Fisher-Privman theory of finite size effects in first order phase transitions \cite{privmann_fisher}.
In the network Ising model constructed from the $2d$ array of boxes and connecting strips, we see that the parameters
can be tuned to produce long-range order, in itself not perhaps surprising, but with
{\it extraordinarily long connecting links}; this diverges exponentially with system  
width, on a scale of the inverse surface tension. Thus, ordering between boxes is feasible over length scales  of many thousands
of molecular diameters.
\acknowledgements
DBA acknowledges with gratitude the support of Prof. S. Dietrich at the MPI Stuttgart, where parts of this work were done.
 

\begin{thebibliography}{99}
%
\bibitem{PKMG} 
J. K. Perron, M. O. Kimball, K. P. Mooney, and F. M. Gasparini, Nature Physics {\bf 6}, 499 (2010).
%
\bibitem{fisher} M. E. Fisher,  Nature Physics {\bf 6}, 483 (2010).
%
\bibitem{PG} 
J. K. Perron and F. M. Gasparini, Phys. Rev. Lett. {\bf 109}, 035302 (2012).
%
\bibitem{O44} L. Onsager,  Phys. Rev. {\bf 65}, 117, (1944); 
R. Peierls, Mathematical Proceedings of the Cambridge Philosophical Society {\bf 32}, 477 (1936). 
%
\bibitem{RCW} C. Ruge, P. Zhu, and F.  Wagner Physica A {\bf 209}, 431 (1994).
%
\bibitem{griffiths} R. B. Griffiths, J. Math. Phys. {\bf 8}, 478 (1967). 
%
\bibitem{privmann_fisher}
V. Privman and M. E.  Fisher, J. Stat. Phys. {\bf 33} 385, (1983).
%
\bibitem{MacCoy_Wu}  B. M. McCoy and  T. T. Wu, Phys. Rev. {\bf 162}, 436 (1967).
%
\bibitem{DBA} D. B. Abraham, Stud. Appl. Math. {\bf 50}, 71 (1971).
%
\bibitem{EPL} D. B. Abraham and A. Macio\l ek, EPL {\bf 101}, 20006 (2013).
%
\bibitem{AM} D. B. Abraham and A.  Macio\l ek, to be published.
%
\bibitem{LB} D. P. Landau and K. Binder, {\it A Guide to Monte Carlo
  Simulations in Statistical Physics} (Cambridge University Press, London,
  2005), p. 155.
%
\bibitem{fisher_au_yang} H.  Au-Yang  and M. E. Fisher, Phys. Rev. E {\bf 88}, 032147 (2013).
%
\bibitem{AMV} D. B. Abraham, A.  Macio\l ek and O. Vasilyev, to be published.
%


\end{thebibliography}
\end{document}